# Building Wavelet Histograms on Large Data in MapReduce

Jeffrey Jestes[1]    Ke Yi[2]    Feifei Li[1]

[1] School of Computing, University of Utah
Salt Lake City, Utah, USA
{jestes,lifeifei}@cs.utah.edu

[2] Dept of Computer Science & Engineering
HKUST, Hong Kong, China
yike@cse.ust.hk

## ABSTRACT

MapReduce is becoming the *de facto* framework for storing and processing massive data, due to its excellent scalability, reliability, and elasticity. In many MapReduce applications, obtaining a compact accurate summary of data is essential. Among various data summarization tools, histograms have proven to be particularly important and useful for summarizing data, and the wavelet histogram is one of the most widely used histograms. In this paper, we investigate the problem of building wavelet histograms efficiently on large datasets in MapReduce. We measure the efficiency of the algorithms by both end-to-end running time and communication cost. We demonstrate straightforward adaptations of existing exact and approximate methods for building wavelet histograms to MapReduce clusters are highly inefficient. To that end, we design new algorithms for computing exact and approximate wavelet histograms and discuss their implementation in MapReduce. We illustrate our techniques in Hadoop, and compare to baseline solutions with extensive experiments performed in a heterogeneous Hadoop cluster of 16 nodes, using large real and synthetic datasets, up to hundreds of gigabytes. The results suggest significant (often orders of magnitude) performance improvement achieved by our new algorithms.

## 1. INTRODUCTION

MapReduce is becoming the *de facto* framework for storing and processing massive data, due to its excellent scalability, reliability, and elasticity [15]. Efficient data processing in MapReduce has received lots of attention since its debut. Development for its open-source realization, Hadoop [22], has been particularly active, e.g., HadoopDB [1], Hadoop++ [16], MapReduce Online [11], Pig [19,29], Hive [36] and others. Datasets stored and processed in Hadoop or any MapReduce platform are usually enormous, ranging from tens of gigabytes to terabytes [15,31]. Hence, in many MapReduce applications, obtaining a compact accurate summary of a dataset is important. Such a summary captures essential statistical properties of the underlying data distribution, and offers quick insight on the gigantic dataset, provided we can compute it efficiently. For example, this allows other MapReduce jobs over the same dataset to better partition the dataset utilizing its histogram which leads to better load-balancing in the MapReduce cluster [15].

In traditional database systems and many modern data management applications, an important useful summary on large datasets is the histogram [32]. Suppose the keys of a dataset are drawn from finite domain $[u] = \{1, \cdots, u\}$. Broadly speaking, a *histogram* on the dataset is any compact, possibly lossy, representation of its frequency vector $\mathbf{v} = (\mathbf{v}(1), \ldots, \mathbf{v}(u))$, where $\mathbf{v}(x)$ is the number of occurrences of key $x$ in the dataset. There are many different histograms depending on the form this compact representation takes. One popular choice is the *wavelet histogram* [26]. Treating $\mathbf{v}$ as a signal, the wavelet histogram consists of the top-$k$ wavelet coefficients of $\mathbf{v}$ in terms of their magnitudes (absolute values), for a parameter $k$. As most real-world distributions have few large wavelet coefficients with others close to zero, retaining only the $k$ largest yields a fairly accurate representation of $\mathbf{v}$. Due to its simplicity, good accuracy, and a variety of applications in data analysis and query optimization, wavelet histograms have been extensively studied. Efficient algorithms are well known for building a wavelet histogram on offline data [26, 27] and for dynamically maintaining it in an online or streaming [13, 20, 27] fashion.

In this work, we study how to efficiently build wavelet histograms for large datasets in MapReduce. We utilize Hadoop to demonstrate our ideas, which should extend to any other MapReduce implementation. We measure the efficiency of all algorithms in terms of *end-to-end running time* (affected by the computation and IO costs) and *intra-cluster communication* (since network bandwidth is also scarce in large data centers running MapReduce [15], whose usage needs to be optimized). Note that communication cost might not be significant when running only one particular MapReduce job (this is often the case); however, in a busy data center/cluster where numerous jobs might be running simultaneously, the aggregated effect from the total communications of these jobs is still critical.

We show straightforward adaptations of both exact and approximate wavelet histogram construction methods from traditional data management systems and data mining fields to MapReduce clusters are highly inefficient, mainly since data is stored in a distributed file system, e.g., the Hadoop Distributed File System (HDFS).

**Contributions.** We propose novel exact and approximation algorithms tailored to MapReduce clusters, in particular Hadoop, which outperform straightforward adaptations of existing methods by several orders of magnitude in performance. Specifically, we:

- present a straightforward adaptation of the exact method in Hadoop, and a new exact method that can be efficiently instantiated in MapReduce in Section 3;

- show how to apply existing, sketch-based approximation algorithms in Hadoop, and discuss their shortcomings. We design a novel random sampling scheme to compute approximate wavelet histograms efficiently in Hadoop in Section 4;

- conduct extensive experiments on large (up to 400GB) datasets in a heterogeneous Hadoop cluster with 16 nodes in Section 5. The experimental results demonstrate convincing results that both our exact and approximation methods have outperformed their counterparts by several orders of magnitude.

In addition, we introduce necessary background on MapReduce, Hadoop, and wavelet histograms in Section 2, survey related work in Section 6, and conclude in Section 7.

## 2. PRELIMINARIES

### 2.1 Wavelet Basics

Suppose each record in the dataset has a key drawn from domain $[u] = \{1, \cdots, u\}$, and we want to build a wavelet histogram on the keys. Define the frequency vector as $\mathbf{v} = (\mathbf{v}(1), \ldots, \mathbf{v}(u))$ where $\mathbf{v}(x)$ is the number of occurrences of key $x$ in the dataset. The idea of building a histogram using wavelets is to consider $\mathbf{v}$ as a signal and apply a wavelet transformation. For most applications, one usually adopts the simplest Haar wavelet basis [13, 18, 20, 26, 27], which is defined as follows. We first average values pairwise to obtain the *average coefficients*, i.e. $[(\mathbf{v}(2) + \mathbf{v}(1))/2, (\mathbf{v}(4) + \mathbf{v}(3))/2, \ldots, (\mathbf{v}(u) + \mathbf{v}(u-1))/2]$. We also retain the average difference of the pairwise values, i.e. $[(\mathbf{v}(2) - \mathbf{v}(1))/2, \ldots, (\mathbf{v}(u) - \mathbf{v}(u-1))/2]$, which are called the *detail coefficients*. Clearly, given these vectors one can reconstruct the original signal $\mathbf{v}$ exactly. We recursively apply this pairwise averaging and differencing process on the *average coefficients vector* until we reach the *overall average* for $\mathbf{v}$. The Haar *wavelet coefficients* of $\mathbf{v}$ are given by the overall average, followed by the detail coefficients in a binary tree, as shown by example in Figure 1, where the leaf level of the tree (level $\ell = \log u$) is the original signal. To preserve the energy of the signal ($\mathbf{v}$'s $L_2$ norm), one must multiply coefficients in level $\ell$ by a scaling factor $\sqrt{u/2^\ell}$.

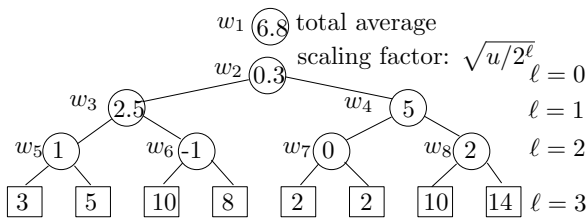

**Figure 1: Wavelet coefficients.**

This transformation is lossless as we can reconstruct $\mathbf{v}$ exactly from all $u$ wavelet coefficients. However, the main reason wavelets are popular and powerful in signal processing is, for most real-world signals $\mathbf{v}$, most of its wavelet coefficients are near zero. Thus if for a parameter $k$ we keep only the $k$ wavelet coefficients of largest magnitude while assuming others are zero, we can still reconstruct the original signal reasonably well. Since *energy* is preserved under wavelet transform, i.e., $\|\mathbf{v}\|_2^2 = \sum_{i=1}^{u} \mathbf{v}(i)^2 = \sum_{i=1}^{u} w_i^2$, keeping the $k$ wavelet coefficients of largest magnitude minimizes energy loss for all $k$-term wavelet representations of $\mathbf{v}$. The best $k$-term wavelet representation can be computed efficiently in a centralized setting [26]: Assuming entries in frequency vector $\mathbf{v}$ are given in order, one can compute all wavelet coefficients bottom-up in $O(u)$ time. Then, using a priority queue of size $k$, we can find the $k$ coefficients of largest magnitude in one pass over all $u$ coefficients, taking time $O(u \log k)$.

| $\frac{(\mathbf{v}(1)+\mathbf{v}(2)+\mathbf{v}(3)+\mathbf{v}(4)+\mathbf{v}(5)+\mathbf{v}(6)+\mathbf{v}(7)+\mathbf{v}(8))}{2\sqrt{2}}$ | | | | | | | |
|---|---|---|---|---|---|---|---|
| $\frac{(\mathbf{v}(5)+\mathbf{v}(6)+\mathbf{v}(7)+\mathbf{v}(8))-(\mathbf{v}(1)+\mathbf{v}(2)+\mathbf{v}(3)+\mathbf{v}(4))}{2\sqrt{2}}$ | | | | | | | |
| $\frac{(\mathbf{v}(3)+\mathbf{v}(4))-(\mathbf{v}(1)+\mathbf{v}(2))}{2}$ | | | | $\frac{(\mathbf{v}(7)+\mathbf{v}(8))-(\mathbf{v}(5)+\mathbf{v}(6))}{2}$ | | | |
| $\frac{\mathbf{v}(2)-\mathbf{v}(1)}{\sqrt{2}}$ | | $\frac{\mathbf{v}(4)-\mathbf{v}(3)}{\sqrt{2}}$ | | $\frac{\mathbf{v}(6)-\mathbf{v}(5)}{\sqrt{2}}$ | | $\frac{\mathbf{v}(8)-\mathbf{v}(7)}{\sqrt{2}}$ | |
| $\mathbf{v}(1)$ | $\mathbf{v}(2)$ | $\mathbf{v}(3)$ | $\mathbf{v}(4)$ | $\mathbf{v}(5)$ | $\mathbf{v}(6)$ | $\mathbf{v}(7)$ | $\mathbf{v}(8)$ |

**Figure 2: Coefficients by wavelet basis vectors**

Another method to compute wavelet coefficients, especially in streaming settings, is to use wavelet basis vectors. The first wavelet basis vector is $\psi_1 = [1, \ldots, 1]/\sqrt{u}$. To define the other $u - 1$ basis vectors, we first introduce, for $j = 1, \ldots, \log u$ and $k = 0, \ldots, 2^j - 1$, the vector $\phi_{j,k}(l) = 1$ for $k(u/2^j) + 1 \leq l \leq k(u/2^j) + u/2^j$, and 0 elsewhere. For $j = 0, \ldots, \log u - 1$ and $k = 0, \ldots, 2^j - 1$, we define the $i$-th wavelet basis vector for $i = 2^j + k + 1$ as $\psi_i = (-\phi_{j+1,2k} + \phi_{j+1,2k+1})/\sqrt{u/2^j}$, where $\sqrt{u/2^j}$ is a scaling factor. The wavelet coefficients are the dot products of $\mathbf{v}$ with these wavelet basis vectors, i.e., $w_i = \langle \mathbf{v}, \psi_i \rangle$, for $i = 1, \ldots, u$; see Figure 2 for an illustration.

Wavelets provide a compact approximation of a data distribution. It serves a variety of data analysis tasks such as range selectivity estimation [26], approximating queries [9] and many other data mining applications [3, 21, 33]. As we are concerned with constructing a best $k$-term wavelet representation, we will not talk about its use which has already been well studied [26].

Wavelet histograms also extend to multi-dimensional signals or datasets. Consider the two-dimensional case where keys are drawn from two-dimensional domain $[u]^2$, defining a two-dimensional frequency array $\mathbf{v} = (\mathbf{v}(x, y)), 1 \leq x, y \leq u$. A 2D wavelet transform first applies a standard 1D wavelet transform to each row of $\mathbf{v}$. Then, using the 1D wavelet coefficients as inputs, we apply a second round of 1D wavelet transforms to each column of the array. This process can be similarly extended to $d$ dimensions.

### 2.2 Hadoop Basics

For this work we assume Hadoop's default file system HDFS. A cluster using HDFS consists of multiple DataNodes, for storing file system data, and a single master node designated as the NameNode which oversees all file operations and maintains all file meta-data. A file in HDFS is split into *data chunks*, 64MB in size by default, which are allocated to DataNodes by the NameNode. Chunks are typically replicated to multiple DataNodes, based on the file replication ratio, to increase data availability and fault tolerance. In this work and many other studies where fault tolerance is not the main subject of interest, the replication ratio is set to 1 and machine failure is not considered. The MapReduce core consists of one master JobTracker task and many TaskTracker tasks. Typical configurations run the JobTracker and NameNode on the same machine, called the *master*, and run TaskTracker and DataNode tasks on other machines, called *slaves*.

Typical MapReduce jobs consist of three phases: Map, Sort-and-Shuffle, and Reduce. The user may specify $m$, the desired number of Mapper tasks, and $r$, the number of Reducer tasks before starting the job. Next we look at the three phases in detail.

**Map phase.** In the Map phase, the $m$ Mappers run in parallel on different TaskTrackers over different logical portions of an input file, called *splits*. Splits typically, but not always, correspond to physical data chunks. Hadoop allows users to specify the InputFormat for a file, which determines how splits are created and defines a RecordReader for reading data from a split.

After splits have been formed, the JobTracker assigns each available Mapper a split to process. By default, the scheduler attempts to schedule Data-Local Mappers by assigning a Mapper a locally stored split. There are also cases which call for Non-Data-Local Mappers, i.e., when a node is idle and has no local split to process. Then, a MapRunner is started which obtains a RecordReader and invokes the Map function for each record in the split. A Mapper then maps input key-value pairs $(k_1, v_1)$ from its split to intermediate key-value pairs $(k_2, v_2)$. As a Mapper proceeds, it maintains an in-memory buffer of the $(k_2, v_2)$. When the buffer fills to threshold, pairs are partitioned, sorted, and optionally processed by the Combine function, which outputs locally aggregated $(k_2, v_2)$ pairs (aggregation on $v_2$'s with the same key $k_2$). Pairs are then spilled to their corresponding logical partitions on the local disk. The partitions are defined by a Partition function, typically a hash function like $hash(k_2) \mod r$, which determines the Reducer task which will process a particular $k_2$ later. When the Mapper ends, all emitted $(k_2, v_2)$ have been partitioned, sorted (w.r.t. $k_2$), and optionally combined. One can also define a Close interface which executes at the end of the Mapper.

**Shuffle-and-Sort phase.** In the Shuffle-and-Sort Phase, each Reducer copies all $(k_2, v_2)$ it is responsible for (as designated by the Partition function) from all DataNodes. It then sorts all received $(k_2, v_2)$ by $k_2$ so all occurrences of key $k_2$ are grouped together. An external sort is needed if the $(k_2, v_2)$ do not fit in memory.

**Reduce phase.** After all $(k_2, v_2)$ are collected and sorted, a Reducer iterates over all its $(k_2, v_2)$. For each distinct key $k_2$, the Reducer passes all corresponding $v_2$ values to the Reduce function. Then the Reduce function produces a final key-value pair $(k_3, v_3)$ for every intermediate key $k_2$. As in the Map phase, one can implement a Close interface which is executed at the end of the Reducer.

## 3. EXACT COMPUTATION

**Baseline solutions** Let $n$ be the total number of records in the entire dataset, where each record has a key drawn from key domain $[u]$. Note either $n \gg u$ or $n \ll u$ is possible. Recall in Hadoop the $n$ records are partitioned into $m$ splits, processed by $m$ Mappers, possibly on different machines, which emit intermediate key-value pairs for processing by Reducers. Thus, one baseline solution to compute the wavelet representation is to compute, for each split $j = 1, \ldots, m$, its local frequency vector $\mathbf{v}_j$, and emit a $(x, \mathbf{v}_j(x))$ pair for each key $x$ in the split. The Reducer then can aggregate the local frequencies, producing the overall frequency vector $\mathbf{v} = \sum_{j=1}^{m} \mathbf{v}_j$ where $\mathbf{v}_j(x) = 0$ if key $x$ does not appear in the $j$th split. Finally, we compute the best $k$-term wavelet representation of $\mathbf{v}$ using the centralized algorithm (e.g., [26]).

We observe that each wavelet coefficient $w_i = \langle \mathbf{v}, \psi_i \rangle$ can be written as

$$w_i = \left\langle \sum_{j=1}^{m} \mathbf{v}_j, \psi_i \right\rangle = \sum_{j=1}^{m} \langle \mathbf{v}_j, \psi_i \rangle,$$

i.e., $w_i$ is the summation of the corresponding local wavelet coefficients of frequency vectors for the $m$ splits. Then, an alternate approach to compute the exact wavelet coefficients is to compute, for each split $j = 1, \ldots, m$ its local frequency vector $\mathbf{v}_j$. The local coefficients $w_{i,j} = \langle \mathbf{v}_j, \psi_i \rangle$ are computed for each split's local frequency vector $\mathbf{v}_j$ and a $(i, w_{i,j})$ pair is emitted for each non-zero $w_{i,j}$. The Reducer can then determine the exact $w_i$ as $\sum_{j=1}^{m} w_{i,j}$ where $w_{i,j} = 0$ if the Reducer does not receive a $w_{i,j}$ from the $j$th split. After computing all complete $w_i$ the Reducer selects the best $k$-term wavelet representation, i.e. by selecting the top-$k$ coefficients of largest absolute value.

A big drawback of the baseline solutions is they generate too many intermediate key-value pairs, $O(mu)$ of them to be precise. This consumes too much network bandwidth, which is a scarce resource in large data clusters shared by many MapReduce jobs [15].

**A new algorithm.** Since $w_i$ is the summation of the corresponding local wavelet coefficients of frequency vectors for the $m$ splits, if we first compute the local coefficients $w_{i,j} = \langle \mathbf{v}_j, \psi_i \rangle$, the problem is essentially a distributed top-$k$ problem. A major difference is in the standard distributed top-$k$ problem, all local "scores" are non-negative, while in our case, wavelet coefficients can be *positive and negative*, and we want to find the top-$k$ aggregated coefficients of *largest absolute value* (magnitude). Negative scores and finding largest absolute values are a problem for existing top-$k$ algorithms such as TPUT and others [7, 17, 28], as they use a "partial sum" to prune items which cannot be in the top-$k$. That is, if we have seen (at the coordinator) $t$ local scores for an item out of $m$ total local scores, we compute a partial sum for it assuming its other $m - t$ scores are zero. When we see $k$ such partial sums we use the $k$th largest partial sum as a threshold, denoted $\tau$, to prune other items: If an item's local score is always below $\tau/m$ at all sites, it can be pruned as it cannot get a total score larger than $\tau$ to get in the top-$k$. If there are negative scores and when the goal is to find largest absolute values, we cannot compute such a threshold as unseen scores may be very negative.

We next present a distributed algorithm which handles positive and negative scores (coefficients) and returns the top-$k$ aggregated scores of largest magnitude. The algorithm is based on algorithm TPUT [7], and can be seen as interleaving two instances of TPUT. As TPUT, Our algorithm requires three rounds. For an item $x$, $r(x)$ denotes its aggregated score and $r_j(x)$ is its score at node $j$.

**Round 1:** Each node first emits the $k$ highest and $k$ lowest (i.e., most negative) scored items. For each item $x$ seen at the coordinator, we compute a lower bound $\tau(x)$ on its total score's magnitude $|r(x)|$ (i.e., $|r(x)| \geq \tau(x)$), as follows. We first compute an upper bound $\tau^+(x)$ and a lower bound $\tau^-(x)$ on its total score $r(x)$ (i.e., $\tau^-(x) \leq r(x) \leq \tau^+(x)$): If a node sends out the score of $x$, we add its exact score. Otherwise, for $\tau^+(x)$, we add the $k$-th highest score this node sends out and for $\tau^-(x)$ we add the $k$-th lowest score. Then we set $\tau(x) = 0$ if $\tau^+(x)$ and $\tau^-(x)$ have different signs and $\tau(x) = \min\{|\tau^+(x)|, |\tau^-(x)|\}$ otherwise. Doing so ensures $\tau^-(x) \leq r(x) \leq \tau^+(x)$ and $|r(x)| \geq \tau(x)$.

Now, we pick the $k$-th largest $\tau(x)$, denoted as $T_1$. This is a threshold for the magnitude of the top-$k$ items.

**Round 2:** A node $j$ next emits all local items $x$ having $|r_j(x)| > T_1/m$. This ensures an item in the true top-$k$ in magnitude must be sent by at least one node after this round, because if an item is not sent, its aggregated score's magnitude can be no higher than $T_1$.

Now, with more scores available from each node, we refine the upper and lower bounds $\tau^+(x), \tau^-(x)$, hence $\tau(x)$, as previously for each item $x \in R$, where $R$ is the set of items ever received. If a node did not send the score for some $x$, we can now use $T_1/m$ (resp. $-T_1/m$) for computing $\tau^+(x)$ (resp. $\tau^-(x)$). This produces a new better threshold, $T_2$ (calculated in the same way as computing $T_1$ with improved $\tau(x)$'s), on the top-$k$ items' magnitude.

Next, we further prune items from $R$. For any $x \in R$ we compute its new threshold $\tau'(x) = \max\{|\tau^+(x)|, |\tau^-(x)|\}$ based on refined upper and lower bounds $\tau^+(x), \tau^-(x)$. We delete item $x$ from $R$ if $\tau'(x) < T_2$. The final top-$k$ items must be in the set $R$.

**Round 3:** Finally, we ask each node for the scores of all items in $R$. Then we compute the aggregated scores exactly for these items, from which we pick the $k$ items of largest magnitude.

A simple optimization is, in Round 2, to not send an item's local score if it is in the local top-$k$/bottom-$k$ sets, even if $|r_j(x)| >$

$T_1/m$; these scores were sent in Round 1. Also in Round 3, a node can send an item's local score only if it was not sent to the coordinator in previous rounds (using simple local bookkeeping).

**Multi-dimensional wavelets.** It is straightforward to extend our algorithms to build multi-dimensional wavelet histograms. Consider the two-dimensional case. Recall in this case frequency vector **v** is a 2D array. A 2D wavelet transform applies two rounds of 1D wavelet transforms on the rows and then the columns of **v**. Since each wavelet transform is a linear transformation, the resulting 2D wavelet coefficients are still linear transformations of **v**. So if we apply a 2D wavelet transform to each split, any 2D wavelet coefficient is still a summation of corresponding 2D coefficients of all splits. Thus, we can still run the modified TPUT algorithm to find the top-$k$ coefficients of largest magnitude as before.

**System issues.** At a high-level, for a dataset in HDFS with $m$ splits, we assign one Mapper task per split and each Mapper acts as a distributed node. We use one Reducer as the coordinator. We implement our three-round algorithm in three rounds of MapReduce in Hadoop. To be consistent across rounds, we identify each split with its unique offset in the original input file.

Two technical issues must be dealt with when implementing the algorithm in Hadoop. First, the algorithm is designed assuming the coordinator and distributed nodes are capable of bi-directional communication. However, in MapReduce data normally flows in one direction, from Mappers to Reducers. In order to have two-way communication we utilize two Hadoop features: the Job Configuration and Distributed Cache. The Job Configuration is a small piece of information communicated to every Mapper and Reducer task during task initialization. It contains some global configuration variables for Mappers and Reducers. The Job Configuration is good for communicating a small amount of information. If large amounts of data must be communicated, we use Hadoop's Distributed Cache feature. A file can be submitted to the master for placement into the Distributed Cache. Then, Distributed Cache content is replicated to all slaves during the MapReduce job initialization.

Second, the distributed nodes and the coordinator in the algorithm need to keep persistent state across three rounds. To do so, at the end of a Mapper task handling an input split, via its Close interface, we write all necessary state information to an HDFS file with a file name identifiable by the split's id. When this split is assigned to a Mapper in a subsequent round, the Mapper can then restore the state information from the file. Note Hadoop always tries to write an HDFS file locally if possible, i.e., state information is usually saved on the same machine holding the split, so saving state information in an HDFS file incurs almost no extra communication cost. For the Reducer which acts as the coordinator, since there is no split associated to it, we choose to customize the JobTracker scheduler so the Reducer is always executed on a designated machine. Thus, the coordinator's state information is saved locally on this machine.

We detail how we address these challenges in Appendix A.

## 4. APPROXIMATE COMPUTATION

We observe the exact computation of the best $k$-term wavelet representation in Hadoop is expensive. Although our improved algorithm avoids emitting all local frequency vectors, it could still be expensive due to the following: (1) The (modified) TPUT algorithm could still send out a lot of communication, though better than sending all local frequency vectors; (2) it needs 3 rounds of MapReduce, which incurs a lot of overhead; and (3) most importantly, every split needs to be scanned to compute local frequency vector $\mathbf{v}_j$ and compute local wavelet coefficients $w_{i,j}$. This motivates us to explore approximation algorithms which compute a $k$-term wavelet representation which may not be the best one, but still approximates the underlying data distribution reasonably well.

There are many design choices for approximate computation of wavelets. Here are some natural attempts: (i) We can replace TPUT with an approximate top-$k$ algorithm [28, 30], after appropriate modification to handle negative scores. This resolves issue (1) but not (2) and (3). (ii) We can approximate local wavelet coefficients of each split using a sketch as in [13, 20], and then send out and combine the sketches, due to the property that these sketches are linearly combinable. This resolves issues (1) and (2), but not (3), as computing a sketch still needs to scan the data once. (iii) Lastly, a generic approach is random sampling, that is, we take a random sample of the keys and construct the wavelets on the sample, as the sample approximates the underlying data distribution well for a sufficiently large sample size. Then a wavelet representation can be constructed on the frequency vector of the sample.

Among the possibilities, only (iii) resolves all three issues simultaneously. It requires only one round, clearing issue (2). It also avoids reading the entire data set, clearing issue (3). However, it may result in a lot of communication, as it is well known to approximate each (global) frequency $\mathbf{v}(x)$ with a standard deviation of $\varepsilon n$ (recall $n$ is the number of records in the entire dataset), a sample of size $\Theta(1/\varepsilon^2)$ is required [37]. More precisely, for a sample probability $p = 1/(\varepsilon^2 n)$ (a sample of expected size $pn = 1/\varepsilon^2$), one can show $\widehat{\mathbf{v}}(x) = \mathbf{s}(x)/p$ is an unbiased estimator of $\mathbf{v}(x)$ with standard deviation $O(\varepsilon n)$ for any $x$, where **s** is the frequency vector of the sample. After that, we construct a wavelet representation on the estimated frequency vector $\widehat{\mathbf{v}}$. As $n$ is the size of the entire data set, which is usually extremely large (for MapReduce clusters), $\varepsilon$ needs to be fairly small for $\widehat{\mathbf{v}}$ to approximate **v** well, usually on the order of $10^{-4}$ to $10^{-6}$. The total communication cost of this *basic sampling* method is $O(1/\varepsilon^2)$, even with one-byte keys, this corresponds to 100MB to 1TB of data being emitted to the network!

A straightforward improvement is to summarize the sampled keys of a split before emitting them, which is actually used as a simple optimization for executing any MapReduce job [15]. We aggregate the keys with the Combine function, that is, if the split is emitting $c$ pairs $(x, 1)$ for the same key, they are aggregated as one pair $(x, c)$. This optimization indeed reduces communication cost, but its effectiveness highly depends on the data distribution, in the worst case it may not reduce the communication at all.

A slightly better idea is to ignore those sampled keys with low frequencies in a split, which we denote as the *improved sampling* algorithm. More precisely, we only send out a sampled key $x$ and its sampled count $\mathbf{s}_j(x)$ if $\mathbf{s}_j(x) \geq \varepsilon t_j$, where $t_j$ is the total number of sampled records in split $j$. Thus the overall error in the total count of a sampled key $x$ from all splits is at most $\sum_{j=1}^{m} \varepsilon t_j = \varepsilon pn = 1/\varepsilon$, which translates into an $(1/\varepsilon)/p = \varepsilon n$ error in the estimated frequency $\widehat{\mathbf{v}}(x)$. Thus it adds another $\varepsilon n$ to the standard deviation, which is still $O(\varepsilon n)$. Note that the total number of key-value pairs sent out by one split is at most $t_j/(\varepsilon t_j) = 1/\varepsilon$. Hence, the total communication of this approach is at most $O(m/\varepsilon)$, which improves upon sending all the samples since usually we have $m \ll 1/\varepsilon$. However, an undesired consequence is that the estimator $\widehat{\mathbf{v}}(x)$ will not be unbiased any more: $\mathbf{E}[\widehat{\mathbf{v}}(x)]$ could be $\varepsilon n$ away from $\mathbf{v}(x)$, since this method ignores all the small sample counts $\mathbf{s}_j(x) < \varepsilon t_j$.

Below we detail a new, *two-level sampling* idea, which produces an unbiased estimator $\widehat{\mathbf{v}}(x)$ for $\mathbf{v}(x)$ with standard deviation $O(\varepsilon n)$ as in the basic random sampling algorithm, while improving communication cost to $O(\sqrt{m}/\varepsilon)$. The idea is to obtain an unbiased estimator $\widehat{\mathbf{s}}(x)$ of $\mathbf{s}(x)$, instead of sending all $\mathbf{s}_j(x)$'s to compute

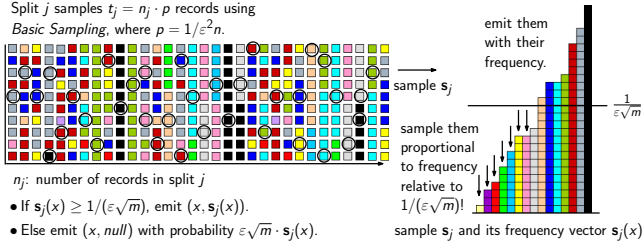
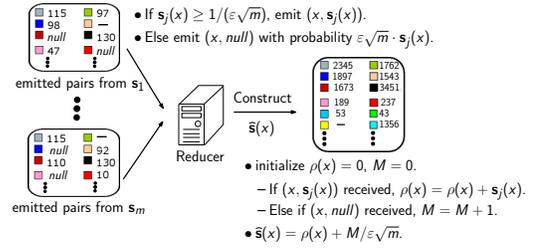

Figure 3: Two-level sampling at mapper.

Figure 4: Two-level sampling at reducer.

$\mathbf{s}(x)$ exactly. We then use $\widehat{\mathbf{s}}(x)$ to produce $\widehat{\mathbf{v}}(x)$. We perform another level of sampling on the local frequency vector $\mathbf{s}_j$ of sampled keys for each split $j$. Specifically, we sample each key $x$ in the sample with probability $\min\{\varepsilon\sqrt{m} \cdot \mathbf{s}_j(x), 1\}$. More precisely, for any $x$ with $\mathbf{s}_j(x) \geq 1/(\varepsilon\sqrt{m})$, we emit the pair $(x, \mathbf{s}_j(x))$; for any $x$ with $0 < \mathbf{s}_j(x) < 1/(\varepsilon\sqrt{m})$, we sample it with a probability proportional to $\mathbf{s}_j(x)$, i.e., $\varepsilon\sqrt{m} \cdot \mathbf{s}_j(x)$, and emit the pair $(x, $ NULL$)$ if it is sampled (for an example please see Figure 3). Note that in two-level sampling we do not throw away sampled items with small frequencies completely, as what is done in the *Improved sampling* method. Rather, these items are still given a chance to survive in the second-level sample, by sampling them proportional to their frequencies relative to the threshold $1/\varepsilon\sqrt{m}$ (which is established from our analysis below).

Next, we show how to construct from the emitted pairs from all splits, an unbiased estimator $\widehat{\mathbf{s}}(x)$ of $\mathbf{s}(x)$ for any key $x \in [u]$ with standard deviation at most $1/\varepsilon$. As $\mathbf{s}(x) = \sum_{j=1}^m \mathbf{s}_j(x)$, we add up all sample count $(x, \mathbf{s}_j(x))$ pairs received for $x$. They do not introduce any error, and we denote this partial sum as $\rho(x)$. If a split has $\mathbf{s}_j(x) < 1/(\varepsilon\sqrt{m})$, it will not emit $\mathbf{s}_j(x)$, but simply emit $(x, $ NULL$)$ if it is sampled. Suppose we receive $M$ such pairs for $x$. Then our estimator is

$$\widehat{\mathbf{s}}(x) = \rho(x) + M/(\varepsilon\sqrt{m}) \qquad (1)$$

(for an example of how we compute $\widehat{\mathbf{s}}(x)$ at the reducer please see Figure 4).

**Theorem 1** $\widehat{\mathbf{s}}(x)$ *is an unbiased estimator of* $\mathbf{s}(x)$ *with standard deviation at most* $1/\varepsilon$.

PROOF. Without loss of generality, assume in the first $m'$ splits $\mathbf{s}_j(x) < 1/(\varepsilon\sqrt{m})$. Write $M$ as $M = \sum_{j=1}^{m'} X_j$ where $X_j = 1$ if $x$ is sampled in split $j$ and 0 otherwise. Each $X_j$ is an independent Bernoulli trial, so

$$\mathbf{E}[X_j] = \varepsilon\sqrt{m} \cdot \mathbf{s}_j(x), \text{ and}$$

$$\mathbf{Var}[X_j] = \varepsilon\sqrt{m} \cdot \mathbf{s}_j(x)(1 - \varepsilon\sqrt{m} \cdot \mathbf{s}_j(x)) \leq \varepsilon\sqrt{m} \cdot \mathbf{s}_j(x). \quad (2)$$

Thus we have

$$\mathbf{E}[M] = \sum_{j=1}^{m'} \varepsilon\sqrt{m} \cdot \mathbf{s}_j(x) = \varepsilon\sqrt{m}(\mathbf{s}(x) - \rho(x)), \quad (3)$$

i.e., $\mathbf{E}[\widehat{\mathbf{s}}(x)] = \mathbf{s}(x)$ combining (1) and (3).

Next, from (2), we have

$$\mathbf{Var}[M] = \sum_{j=1}^{m'} \mathbf{Var}[X_j] = \varepsilon\sqrt{m} \cdot \sum_{j=1}^{m'} \mathbf{s}_j(x). \quad (4)$$

Since each $\mathbf{s}_j(x) \leq 1/(\varepsilon\sqrt{m})$, $\mathbf{Var}[M]$ is at most $m'$. Thus, the variance of $\widehat{\mathbf{s}}(x)$ is $\mathbf{Var}[M/(\varepsilon\sqrt{m})] = \mathbf{Var}[M]/(\varepsilon^2 m)$. So $\mathbf{Var}[\widehat{\mathbf{s}}(x)] \leq m'/(\varepsilon^2 m) \leq 1/\varepsilon^2$, namely, the standard deviation is at most $1/\varepsilon$. □

From $\widehat{\mathbf{s}}(x)$, we can estimate $\mathbf{v}(x)$ as $\widehat{\mathbf{v}}(x) = \widehat{\mathbf{s}}(x)/p$ (recall that $p = 1/(\varepsilon^2 n)$ is the sampling probability of the first level random sample in each split). It will be an unbiased estimator of $\mathbf{v}(x)$ with standard deviation $(1/\varepsilon)/p = \varepsilon n$.

**Corollary 1** $\widehat{\mathbf{v}}(x)$ *is an unbiased estimator of* $\mathbf{v}(x)$ *with standard deviation at most* $\varepsilon n$.

Corollary 1 gives a bound on the error of the estimated frequencies. Below we also analyze the error in the computed wavelet coefficients. Consider the coefficient $w_i = \langle \mathbf{v}, \psi_i \rangle$, where $\psi_i = (-\phi_{j+1,2k} + \phi_{j+1,2k+1})/\sqrt{u/2^j}$ is the corresponding wavelet basis vector (see discussion in Section 2.1). From the estimated frequency vector $\widehat{\mathbf{v}}$, we estimate $w_i$ as $\widehat{w}_i = \langle \widehat{\mathbf{v}}, \psi_i \rangle$. Since $\widehat{\mathbf{v}}(x)$ for every $x$ is unbiased, $\widehat{w}_i$ is also an unbiased estimator of $w_i$. Recall that $\psi_i(x) = -1, +1$ for $x = 2ku/2^{j+1}+1, \ldots, (2k+2)u/2^{j+1}$, so the variance of $\widehat{w}_i$ is

$$\mathbf{Var}[\widehat{w}_i] = \frac{2^j}{u} \sum_{x=2ku/2^{j+1}+1}^{(2k+2)u/2^{j+1}} \mathbf{Var}[\widehat{\mathbf{v}}(x)]$$

$$= \frac{2^j}{u} \sum_{x=2ku/2^{j+1}+1}^{(2k+2)u/2^{j+1}} \mathbf{Var}[\widehat{\mathbf{s}}(x)]/p^2$$

$$= \frac{2^j}{u} \sum_{x=2ku/2^{j+1}+1}^{(2k+2)u/2^{j+1}} \mathbf{Var}[M]/(\varepsilon^2 m p^2)$$

$$\leq \frac{2^j n}{um} \sum_{x=2ku/2^{j+1}+1}^{(2k+2)u/2^{j+1}} \varepsilon\sqrt{m}\mathbf{s}(x) \qquad \text{(by (4))}$$

$$= \frac{\varepsilon 2^j n}{u\sqrt{m}} \sum_{x=2ku/2^{j+1}+1}^{(2k+2)u/2^{j+1}} \mathbf{s}(x). \qquad (5)$$

Note that $\sum_{x=2ku/2^{j+1}+1}^{(2k+2)u/2^{j+1}} \mathbf{s}(x)$ is just the total number of keys covered by the wavelet basis vector. This discussion leads to the next result:

**Theorem 2** *The two-level sampling method provides an unbiased estimator $\widehat{w}_i$ for any wavelet coefficient $w_i$, and the variance of $\widehat{w}_i$ is bounded by* (5).

Finally, it remains to bound its communication cost.

**Theorem 3** *The expected total communication cost of our two-level sampling algorithm is* $O(\sqrt{m}/\varepsilon)$.

PROOF. The expected total sample size of first-level sampling is $pn = 1/\varepsilon^2$. Thus, there are at most $(1/\varepsilon^2)/(1/(\varepsilon\sqrt{m})) = \sqrt{m}/\varepsilon$ keys with $\mathbf{s}_j(x) \geq 1/(\varepsilon\sqrt{m})$ across all splits. These keys must be emitted for second level sampling. For any key $x$ in any split $j$ with $\mathbf{s}_j(x) < 1/(\varepsilon\sqrt{m})$, we emit it with probability $\varepsilon\sqrt{m} \cdot \mathbf{s}_j(x)$, so the expected total number of sampled keys for this category is

$$\sum_j \sum_x \varepsilon\sqrt{m} \cdot \mathbf{s}_j(x) \le \varepsilon\sqrt{m} \cdot 1/\varepsilon^2 = \sqrt{m}/\varepsilon.$$

So the total number of emitted keys is $O(\sqrt{m}/\varepsilon)$. □

Consider typical values: $m = 10^3, \varepsilon = 10^{-4}$ and 4-byte keys. Basic sampling emits $1/\varepsilon^2 \approx 400$MB; improved sampling emits at most $m/\varepsilon \approx 40$MB; while two-level sampling emits about $\sqrt{m}/\varepsilon \approx 1.2$MB of data—a 330-fold or 33-fold reduction, respectively!

*Remark:* In our second-level sampling, the sampling probability depends on the frequency, so that "important" items are more likely to be sampled. This falls into the general umbrella of "importance sampling" [35], and has been used for frequency estimation on distributed data [23,39]. However, its application to wavelet histograms and the corresponding variance analysis are new.

**Multi-dimensional wavelets.** Our algorithm extends to building multi-dimensional wavelet histograms naturally. In $d$ dimensions, frequency vector $\mathbf{v}$ is a $d$-dimensional array, and frequency array $\mathbf{s}$ of a random sample of the dataset still approximates $\mathbf{v}$. So the problem boils down to how well $\mathbf{s}$ approximates $\mathbf{v}$ (note our two-level sampling algorithm does not affect the approximation error of the sample). However, because data is usually sparse in higher dimensions, the quality of the sample may not be as good as in one dimension. In fact, the standard deviation of the estimated frequency for any $\mathbf{v}(x)$ ($x$ is now a cell in $[u]^d$) from a sample of size $O(1/\varepsilon^2)$ is still $O(\varepsilon n)$, but due to the sparsity of the data, all the $\mathbf{v}(x)$'s may be small, so the relative error becomes larger. This is, unfortunately, an inherent problem with sparse data: if all $\mathbf{v}(x)$'s are small, say 0 or 1, then random sampling, and in general any sublinear method, cannot possibly achieve small relative errors [14]. One remedy is to lower the granularity of the data, i.e., project the data to a smaller grid $[u/t]^d$ for some appropriate $t$ so as to increase the density of the data.

**System issues.** Among the three general approximation strategies mentioned at the beginning of Section 4, implementing the approximate TPUT methods (such as KLEE [28]) in Hadoop requires at least three rounds of MapReduce, which involves too much overhead for just approximating a wavelet histogram. Wavelet sketches can be easily implemented in Hadoop. The idea is to run one Mapper per split, which builds a local wavelet sketch for the split and emits the non-zero entries in the sketch to the Reducer. The Reducer then combines these $m$ sketches and estimates the top-$k$ coefficients from the combined sketch. There are two wavelet sketches in the literature: the AMS sketch [4,20] and the GCS sketch [13]. The latter was shown to have better performance, so we choose it to implement in Hadoop. There are some technical details in optimizing its implementation in Hadoop, which we omit here.

The third strategy, random sampling, clearly has better performance as it avoids scanning the entire dataset and is also easy to implement in Hadoop. Our two-level sampling algorithm in addition achieves very low communication cost. We detail how we address some system issues, overcome the challenges, and implement two-level sampling in Hadoop in Appendix B; implementation of the other sampling algorithms is even simpler.

## 5. EXPERIMENTS

We implement all algorithms in Hadoop and empirically evaluate their performance, in both end-to-end running time and communication cost. For the exact methods, we denote the baseline solution of sending all local frequency vectors (the $\mathbf{v}_j$'s of all splits) in Section 3 as *Send-V*, the baseline solution of sending the local wavelet coefficients (the $w_{i,j}$'s of all splits) in Section 3 as *Send-Coef*, and our new algorithm as *H-WTopk* (meaning "Hadoop wavelet top-$k$"). For the approximate algorithms, we denote the basic sampling method as *Basic-S*, the improved sampling method as *Improved-S*, and the two-level sampling method as *TwoLevel-S*. Note *Improved-S* is based on the same idea as *Basic-S*, but offers strictly better performance, which we derived in Section 4. Given this fact, we choose to utilize *Improved-S* as the default competitor of *TwoLevel-S*. We also implement the sketch-based approximation method as discussed in Section 4. We use the GCS-sketch which is the state-of-the-art sketching technique for wavelet approximations [13]. We denote this method as *Send-Sketch*. We did not attempt to modify the approximate TPUT methods (such as KLEE [28]) to work with negative values and adapt them to MapReduce, since they generally require multiple rounds and scanning the entire datasets, which will be strictly worse than other approximation methods.

**Setup and datasets.** All experiments are performed on a heterogeneous Hadoop cluster running the latest stable version of Hadoop, version 0.20.2. The cluster consists of 16 machines with four different configurations: (1) 9 machines with 2GB of RAM and one Intel Xeon 5120 1.86GHz CPU, (2) 4 machines with 4GB of RAM and one Intel Xeon E5405 2GHz CPU, (3) 2 machines with 6GB of RAM and one Intel Xeon E5506 2.13GHz CPU, and (4) 1 machine with 2GB of RAM and one Intel Core 2 6300 1.86GHz CPU. Our master runs on a machine with configuration (2) and we select one of the machines of configuration (3) to run the (only) Reducer. We configure Hadoop to use 300GB of hard drive space on each slave and allocate 1GB memory per Hadoop daemon. We have one TaskTracker and one DataNode daemon running on each slave, and a single NameNode and JobTracker daemon on the master. All machines are directly connected to a 100Mbps switch.

For the datasets, clearly, the determining parameters are $n$, the total number of records, which corresponds to the size of the input file, and $u$, the domain size, as well as the skewness. Note it only makes sense to use a dataset which is at least tens of gigabytes and has a domain size on the order of $2^{20}$. Otherwise a centralized approach would work just fine, and the overhead of running MapReduce could actually lead to worse performance [15].

That said, for real datasets, we test all algorithms on the *WorldCup* [6] dataset which is the access logs of 92 days from the 1998 World Cup servers, a total of approximately 1.35 billion records. Each record consists of 10 4-byte integer values including month, day, and time of access as well as the client id, object id, size, method, status, and accessed server. We assign to each record a 4-byte identifier *clientobject*, which uniquely identifies a distinct client id and object id pairing. The object id uniquely identifies a URL referencing an object stored on the World Cup servers, such as a page or image. The pairing of the client id and the object id is useful to analyze the correlation between clients and resources from the World Cup servers, under the same motivation as that in the more common example of using the (src ip, dest ip) pairing in a network traffic analysis scenario. There are approximately 400 million distinct client id object id combinations, so the domain of this key value is approximately $2^{29}$, i.e. $u = 2^{29}$. We store *WorldCup* in binary format, and in total the stored dataset is 50GB.

To model the behavior of a broad range of real large datasets, we also generate datasets following the Zipfian distribution (since most real datasets, e.g., the *clientobject* in *WorldCup*, are skewed with different levels of skewness), with various degrees of skewness $\alpha$, as well as different $u$ and $n$. We randomly permute keys in a dataset to ensure the same keys do not appear contiguously in the input file. Each dataset is stored in binary format and contain records with only a 4-byte integer key. Unless otherwise specified, we use the Zipfian dataset as our default dataset to vigorously test all approaches on a variety of parameters on large scale data.

We vary $\alpha$ in $\{0.8, 1.1, 1.4\}$ and $\log_2 u$ in $\{$ 8, 11, 14, 17, 20, 23, 26, 29, 32 $\}$. We vary input file size from 10GB to 200GB resulting in different $n$ from 2.7 to 54 billion. We vary the size of a record from 4-bytes to 100kB. For all algorithms, we use 4-byte integers to represent $\mathbf{v}(x)$ in a Mapper and 8-byte integers in a Reducer. We represent wavelet coefficients and sketch entries as 8-byte doubles.

For all experiments, we vary one parameter while keeping the others fixed at their default values. Our default $\alpha$ is 1.1 and $\log_2 u$ is 29. The default dataset size is 50GB (so the default $n$ is 13.4 billion). The default record size is 4-bytes. We compute the best $k$-term wavelet histogram with $k = 30$ by default, which also varies from 10 to 50. The default split size $\beta$ is 256MB, which varies from 64MB to 512MB. Note that the number of splits is $m = 4n/(1024^2 \beta)$ (so the default $m$ is 200). We also simulate a live MapReduce cluster running in a large data center where typically multiple MapReduce jobs are running at the same time, which share the network bandwidth. Thus, the default available network bandwidth is set to 50% (i.e., 50Mbps) but we also vary it from 10% to 100%. Note, we omit the results for *Send-Coef* on all experiments except for varying the domain $u$ of the input dataset as it performs strictly worse than *Send-V* for other experiments.

The exact methods have no parameters to tune. For *Send-Sketch*, we use a recommended setting for the GCS-sketch from [13], where each sketch is allocated 20KB· $\log_2 u$ space. We use GCS-8 which has the overall best per-item update cost and a reasonable query time to obtain the final coefficients. We also did the following optimizations: First, for each split, we compute the local frequency vector $\mathbf{v}_j$, and then insert the keys into the sketch so we update the sketch only once for each distinct key. Second, we only send non-zero entries in a local sketch to the Reducer. For the two sampling methods, the default $\varepsilon$ is $10^{-4}$, and we vary it from $10^{-5}$ to $10^{-1}$.

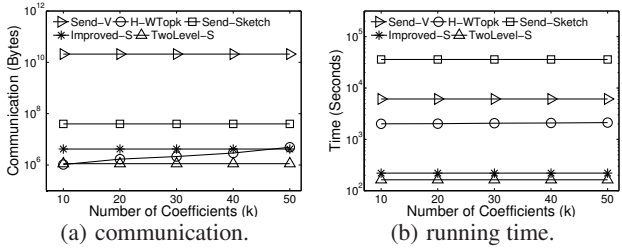

(a) communication.　　(b) running time.

**Figure 5: Cost analysis: vary $k$.**

**Results on varying $k$.** We first study the effect of $k$, i.e., the size of the wavelet histogram to be computed. Figure 5 shows the effect of varying $k$ on the communication cost and running time of all algorithms. The results show $k$ has little impact on performance, except for the communication cost of *H-WTopk*. This is expected, as *Send-V* (resp. *Send-Sketch*) always compute and send out all local frequency vectors (resp. their sketches). The sampling methods are also unaffected by $k$ as the sampling rate is solely determined by $m$ and $\varepsilon$. However, *H-WTopk*'s communication cost is closely related to $k$, as it determines thresholds $T_1$ and $T_2$ for pruning items.

For the exact methods, *H-WTopk* outperforms *Send-V* by orders of magnitude, in both communication and running time. It also outperforms *Send-Sketch*, which is an approximate method. The two sampling algorithms are clearly the overall winners. Nevertheless, among the two, in addition to a shorter running time, *TwoLevel-S* reduces communication to 10%–20% compared to *Improved-S*. Recall our analysis indicates an $O(\sqrt{m})$-factor reduction from

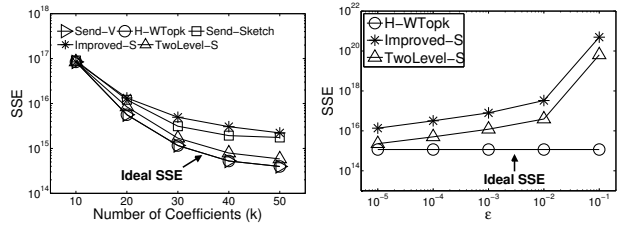

**Figure 6: SSE: vary $k$.**　　**Figure 7: SSE: vary $\varepsilon$.**

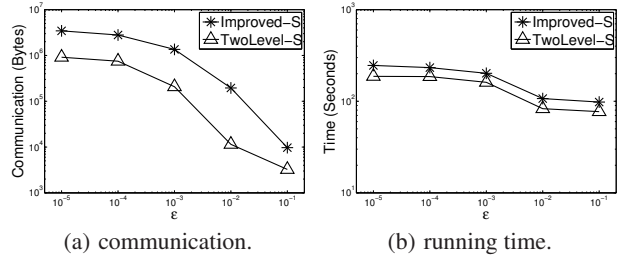

(a) communication.　　(b) running time.

**Figure 8: Cost analysis: vary $\varepsilon$.**

*Improved-S* to *TwoLevel-S*; but this assumes any input data. Due to the skewness of the Zipfian data distribution, *Improved-S* actually combines many keys into one key-value pair, and thus typically does not reach its $O(m/\varepsilon)$ upper bound on communication. Overall, the sampling algorithms have impressive performance: On this 50GB dataset, *TwoLevel-S* incurs only 1MB communication and finishes in less than 3 minutes. In contrast, *Send-Sketch* takes about 10 hours (most time is spent updating local sketches), *Send-V* about 2 hours (mostly busy communicating data), and *H-WTopk* 33 minutes (scanning inputs plus overhead for 3 rounds of MapReduce).

We must ensure the efficiency gain of the sampling methods does not come with a major loss of quality. Thus, we examine the sum of squared error (SSE) between the frequency vector reconstructed from the wavelet histogram and that of the original dataset. The results are shown in Figure 6. Since *Send-V* and *H-WTopk* are exact methods, they represent the best possible reconstruction using any $k$-term wavelet representation. So their curves are identical in Figure 6 and represent the ideal error for measuring the accuracy of the approximation methods. Clearly, when $k$ increases, the SSEs of all methods decrease. Among the three approximation methods, *TwoLevel-S* returns wavelet histograms which come very close to the ideal SSE. *Improved-S* has the worst SSE as it is not an unbiased estimator for $\mathbf{v}$, and the gap from the ideal SSE widens as $k$ gets larger, as it is not good at capturing the details of the frequency vector. *Send-Sketch*'s SSE is between *TwoLevel-S* and *Improved-S*. Even though the SSE looks large in terms of absolute values, it is actually quite small considering the gigantic dataset size. When $k \geq 30$, the SSE is less than 1% of the original dataset's energy.

**Varying $\varepsilon$.** Next we explore the impact of $\varepsilon$ on all sampling methods, by varying it from $10^{-5}$ to $10^{-1}$ in Figure 7. In all cases, *TwoLevel-S* consistently achieves significantly better accuracy than *Improved-S*, as the first is an unbiased estimator of $\mathbf{v}$ while the latter is not. Both methods have larger SSEs when $\varepsilon$ increases, with $\varepsilon = 10^{-4}$ achieving a reasonable balance between the SSE and efficiency (to be shown next), hence it is chosen as the default. Figure 8 shows all sampling methods have higher costs when $\varepsilon$ decreases (from right to left). In all cases, *TwoLevel-S* has significantly lower communication cost than *Improved-S* as seen in Figure 8(a). It also has a lower running time compared to *Improved-S* as shown in Figure 8(b). In a busy data center where network bandwidth is shared by many concurrent jobs, the savings in communication by *TwoLevel-S* will prove to be critical and the gap for the running time will widen even more.

In what follows, we omit the results on SSEs when we vary the other parameters, as they have less impact on the SSEs of various methods, and the relative trends on SSEs for all methods are always similar to those reported in Figures 6 and 7.

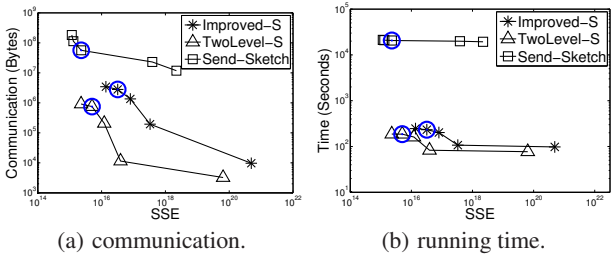

(a) communication.   (b) running time.

**Figure 9: Communication and running time versus SSE.**

**Comparing SSE.** For the next experiment we analyze the communication and computation overheads of all approximation algorithms to achieve a similar SSE in Figure 9, where the defaults of all algorithms are circled. In Figure 9(a) we see that the communication cost increases as the SSE decreases for all algorithms. *TwoLevel-S* achieves the best SSE to communication cost, and communicates at least an order of magnitude less than *Improved-S* and two orders of magnitude less than *Send-Sketch* to achieve a similar SSE. Among the algorithms, *TwoLevel-S* is the most efficient in terms of running time, achieving a similar SSE to *Send-Sketch* in orders of magnitude less time and approximately 2-3 times less time than *Improved-S*, as shown in Figure 9(b). These results also indicate the sketch size selected at $20\text{kB} * \log_2(u)$ is most competitive against the sampling based algorithms, justifying our choice for using it as the default value for the GCS-sketch.

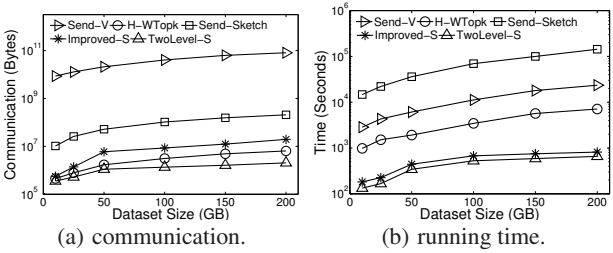

(a) communication.   (b) running time.

**Figure 10: Cost analysis: vary $n$.**

**Varying dataset size $n$.** Next, we analyze the scalability of all methods by varying $n$, or equivalently the dataset size. Note as $n$ increases, so does $m$, the number of splits. This explains the general trends in Figure 10 for both communication and running times. There are two points worth pointing out. First, *TwoLevel-S* outperforms *Improved-S* by a larger margin in terms of communication cost for larger datasets due to the $O(\sqrt{m})$-factor difference, which becomes more than one order of magnitude when the data becomes 200GB. Second, the increase in $m$ leads to longer running times of all methods, but the two sampling algorithms are much less affected. The reason is the sampling algorithms mainly have two kinds of running time costs: overheads associated with processing each split (i.e., Mapper initialization), which linearly depends on $m$, and sampling overheads where the sample size is always $\Theta(1/\varepsilon^2)$, which is independent of $n$. The net effect of these costs is a slow growth in running time. Overall, *H-WTopk* and *TwoLevel-S* are clearly the best exact and approximate methods, respectively.

**Varying record size.** In Figure 11 we analyze the effect varying the record size has on the performance of all algorithms. We fix the number of records as 4,194,304 (which is the number of records when the total dataset reaches 400GB with 100kB per record), for the default Zipfian dataset, and vary the size of a record from 4 bytes (key only) to 100kB, which corresponds to a dataset size of 16MB to 400GB consisting of 1 and 1600 splits respectively. We see the communication cost increases for all methods as the record size increases. This makes sense since increasing the number of splits has a negative impact to all of their communication costs. Nevertheless, even with 1600 splits when the record size is 100kB *H-WTopk* still communicates less than *Send-V*; and *TwoLevel-S* still outperforms the other algorithms by orders of magnitude with respect to communication.

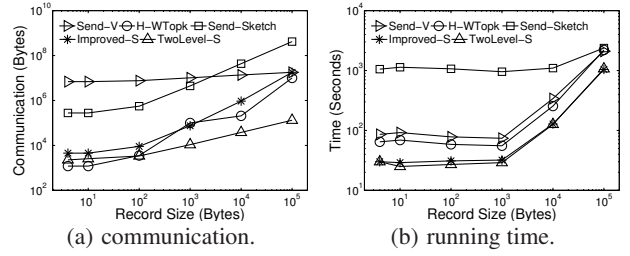

(a) communication.   (b) running time.

**Figure 11: Cost analysis: vary record size.**

The running time of all algorithms also increases as the record size increases, while the total number of records is fixed. This is not surprising due to several factors when the record size increases: 1) all algorithms communicate more data; 2) there are much more splits than the number of slaves in our cluster; 3) the IO cost becomes higher. Note that regardless of the record size *H-WTopk* still performs better than *Send-V*. We also note that the clear winner is *TwoLevel-S* with a running time roughly an order of magnitude better than *Send-V*. Finally, the performance gap between *H-WTopk* and *Send-V*, as well as the gap between *TwoLevel-S* and *Improved-S*, are not as significant as in other experiments. This is mostly due to the small number of records (only 4 million, in contrast to 13.4 billion in the default zipfian dataset and 1.35 billion in the WorldCup dataset) we have to use in this study, which is constrained by the number of records we can accommodate for the maximum record size (100kB), while still keeping the total file size under control (400GB when each record becomes 100kB).

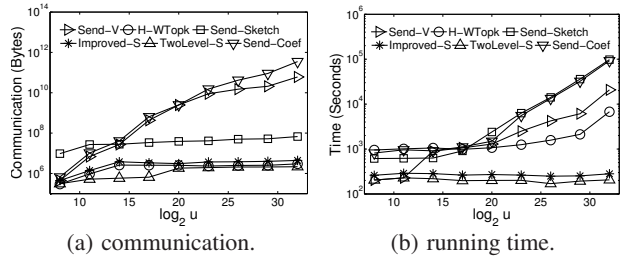

(a) communication.   (b) running time.

**Figure 12: Cost analysis: vary $u$.**

**Varying domain size $u$.** We next examine how $u$ affects all algorithms. Note as we increase $u$ while keeping $n$ fixed, the tail of the data distribution gets longer while frequent elements get slightly less frequent. Figure 12(a) shows this affects *Send-V*, which is obvious as each local frequency vector $\mathbf{v}_j$ gets more entries. We note that *Send-V* performs better than *Send-Coef* for all tested domains. *Send-Coef* reduces the computational burden at the reducer by performing the wavelet transform in parallel over the local frequency vectors. However, the results indicate that the potential savings from computing the wavelet transform in parallel is canceled out by the increase in communication and computation cost of *Send-Coef* over *Send-V*. The overheads in *Send-Coef* are caused by the fact that the number of local wavelet coefficients grows linearly to the domain size, regardless of the size of each split and how many

116

records a local split contains. Thus, with the increasing domain size, the communication cost and the overall running time of this approach quickly degrade. Indeed, the total non-zero local wavelet coefficients are almost always much greater than the total number of keys in the local frequency vector with a non-zero frequency. Since *Send-V* always results in less communication and computation overheads than *Send-Coef*, we use *Send-V* as our default baseline algorithm for all other experiments. In terms of running time, larger $u$ makes all methods slower except the sampling-based algorithms. *Send-V*, *Send-Coef*, *H-WTopk* and *Send-Sketch* all more or less linearly depend on $u$: *Send-V* and *Send-Coef* are obvious; *H-WTopk* needs $O(u)$ time to compute the wavelet transformation for each $\mathbf{v}_j$; while *Send-Sketch* needs to make $O(u)$ updates to the sketch. The two sampling algorithms are not affected as their sample size is independent of $u$.

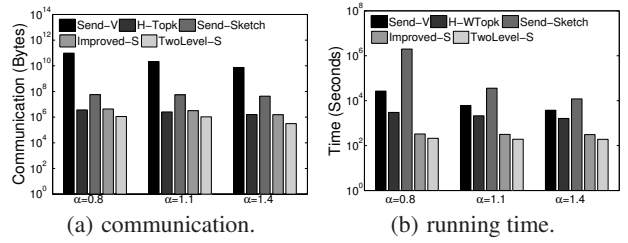

(a) communication. (b) running time.

**Figure 14: Cost analysis: vary skewness $\alpha$.**

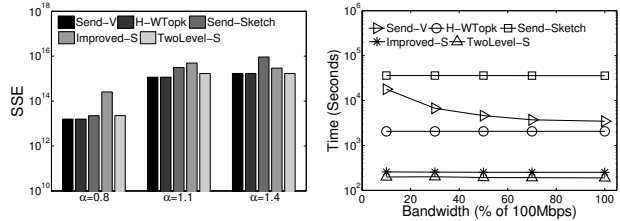

**Figure 15: Vary $\alpha$ SSE.** **Figure 16: Vary B.**

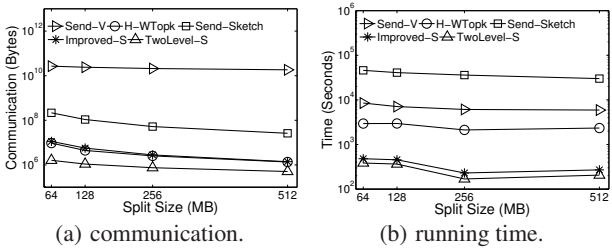

(a) communication. (b) running time.

**Figure 13: Cost analysis: vary split size $\beta$.**

**Varying split size $\beta$.** Figure 13 shows the effect of varying the split size $\beta$ from 64MB to 512MB while keeping $n$ fixed. The number of splits $m$ drops for larger split sizes (varying from 800 to 100 for the 50GB dataset). Hence, the communication cost of all algorithms drop with a larger split size. This is essentially the opposite of Figure 10(a) where we increase $n$ (hence $m$) for a fixed split size. The difference is, for *Send-V*, the communication is not reduced as much, since as the split gets larger, there are more distinct keys in each split, which cancels some benefit of a smaller $m$.

The running times of all methods reduce slightly as well for larger split size, because *Send-V* has less communication overhead, *H-WTopk* has to perform less local wavelet transformations, and *Send-Sketch* has less updates to the local sketches. For the two sampling algorithms, although their sample size does not depend on $m$, the communication (hence the cost of the Reducer who needs to process all the incoming messages) reduces as $m$ gets smaller.

All these seem to suggest we should use a split size as large as possible. However, there is a limit on the split size, constrained by the available local disk space (so that a split does not span over multiple machines, which would incur significant communication cost when processing such a split). In addition, larger split sizes reduce the granularity of scheduling and increase the overhead of failure recovery. On our cluster with 16 machines, these issues do not manifest. But on large clusters with thousands of machines, the split size should not be set too large. So the typical split size as recommended by most works in the literature (e.g. [2, 25, 38]) is either 128MB or 256MB.

**Varying data skewness $\alpha$.** We also study the effect of data skewness $\alpha$, with $\alpha$ as $0.8, 1.1, 1.4$ and show results in Figure 14 and 15. When data is less skewed, each split has more distinct key values. As a result, the communication cost of *Send-V* is higher, leading to higher running time. The running time of *Send-Sketch* becomes more expensive as more local sketch updates are necessary. The communication and running time of other methods have little changes. The SSE is analyzed in Figure 15. All methods' SSE seem to improve on less skewed data. Nevertheless, *TwoLevel-S* consistently performs the best among all approximation methods.

**Varying bandwidth $B$.** Finally, Figure 16 shows the effect the bandwidth $B$ has on the running time of all methods, by varying it from 10% to 100% of the full network bandwidth which is 100Mbps. The communication cost of all algorithms are unaffected by $B$. *Send-V* enjoys an almost linear reduction in running time when $B$ increases as transmitting data dominates its running time. Other methods see a slight reduction in their respective running times.

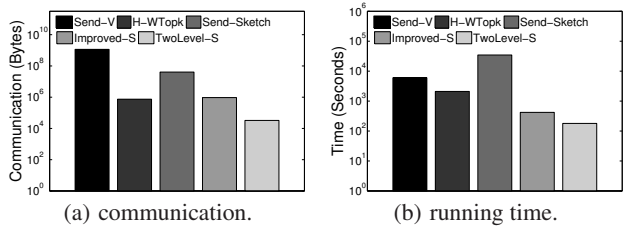

(a) communication. (b) running time.

**Figure 17: Cost analysis: WorldCup dataset.**

**WorldCup Dataset.** Figure 17 analyzes the performance of all algorithms on *WorldCup* using default $k$, $\varepsilon$, $\beta$, and $B$ values, in which we attempt to compute the best $k$-term wavelet representation over the *clientobject* attribute. Notice in Figure 17(a) the communication trends for all algorithms are similar to our previous observations. We note the *WorldCup* dataset is approximately 50GB with almost $2^{29}$ distinct *clientobject* values, which are the defaults used for the Zipfian datasets. *Send-V*'s communication cost is dependent on two primary factors: The skewness of the data and the total number of distinct values. As the data becomes more skewed, *Send-V* can leverage on the Combine function to reduce communication. However, as we see in Figure 17(a) *Send-V* requires roughly the same amount of communication as for the Zipfian datasets. This indicates by varying $\alpha$, $u$ and $n$ for the Zipfian datasets we can approximate the distribution of real large datasets fairly well.

In Figure 17(b) we observe the running times of all approaches on *WorldCup*. *Send-V*'s running time is mainly dependent on its communication cost. The data communicated is about the same as the default Zipfian dataset so it is not surprising *Send-V* preforms similarly on the *WorldCup* dataset. We would like to note *TwoLevel-S* saves almost 2 orders of magnitude and *H-WTopk* saves about 0.5-1 order of magnitude over *Send-V* indicating our algorithms are effective on large real datasets.

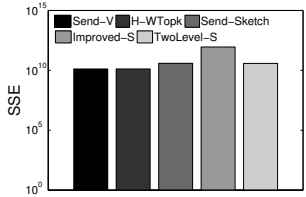

Figure 18: SSE on WorldCup.

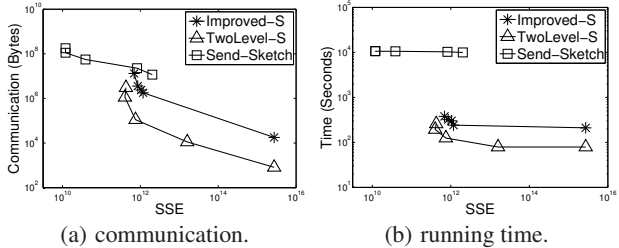

(a) communication.  (b) running time.

Figure 19: Comm. & running time vs SSE on WorldCup.

We observe the SSE on *WorldCup* in Figure 18. The relative performance of various algorithms are similar to the previously observed trends for Zipfian datasets in Figure 15. We also analyze the communication and running time of all algorithms versus the SSE on *WorldCup* in Figure 19. The trends are again similar to that in Figure 9 for Zipfian datasets. Notice the *Send-Sketch* method achieves a similar SSE, with at least an order of magnitude more communication and orders of magnitude more computation overheads than other methods. We observe that *TwoLevel-S* achieves the best overall SSE to communication cost, requiring approximately an order of magnitude less communication than other methods. In addition, *TwoLevel-S* is also 2-3 times or orders of magnitude faster than other methods to achieve a similar SSE.

**Experimental conclusion.** These extensive results reach the clear conclusion *H-WTopk* is the choice if we wish to find exact top-$k$ wavelet coefficients, outperforming the baseline exact method *Send-V* by several orders of magnitude in communication, and 0.5-1 order of magnitude in running time; when approximation is allowed, *TwoLevel-S* is the best method. Not only does it offer the cleanest solution, but it also achieves an SSE nearly as good as exact methods for a tiny fraction of their communication cost and running time. In addition, it achieves the best overall communication and running time to achieve an SSE similar to other sampling and sketching techniques. It produces an approximate wavelet histogram of high approximation quality for 200GB data of domain size of $2^{29}$ in less than 10 minutes with only 2MB communication!

## 6. RELATED WORK

The wavelet histogram and wavelet analysis, introduced to data management for selectivity estimation by Matias et al. [26], has quickly emerged as a widely used tool in databases, data mining, and data analysis [3, 9, 21, 33]. Matias et al. have also studied how to dynamically maintain the wavelet histograms under updates [27]. Gilbert et al. [20] extended the construction of the wavelet histogram to streaming data, using the AMS sketch [4]. Cormode et al. [13] then improved the efficiency of the sketch with the Group-Count Sketch (GCS).

Many types of histograms exist. Poosala et al. [32] presented an excellent discussion on the properties of various histograms. How to efficiently build other types of histograms for large data in MapReduce is an intriguing open problem we plan to investigate.

Since its introduction [15], MapReduce has quickly become a primary framework for processing massive data. It represents the trend of going towards parallel and distributed processing on shared-nothing commodity clusters [8, 10, 31]. Significant effort has been devoted to improving the efficiency, the functionality and query processing in MapReduce, e.g., Amazon EC2 [5], HadoopDB [1], Hadoop++ [16], Hadoop [22], MapReduce Online [11], and many others [12]. To the best of our knowledge, efficient construction of wavelet histograms in MapReduce has not been studied.

Our work is also related to finding distributed top-$k$ and frequent items. The best exact method for distributed top-$k$ is TPUT [7]. However, it (and other methods, e.g., [28]) does not support finding the aggregates with the largest absolute values over positive and negative value sets. Our exact algorithm shares the basic principle in distributed query processing, however, comes with novel designs in order to work for wavelets in MapReduce. The approximate distributed top-$k$ query has been studied in [28] and many others. However, they also only support non-negative scores and require multiple rounds, which introduce considerable overhead in the MapReduce framework. As such, we did not attempt to adapt them as approximation methods. Instead, for approximation methods, we focus on algorithms that require only one round of MapReduce. The most related works are methods for finding heavy hitters from distributed datasets [39]. But they are not tailored for the MapReduce environment, and use complicated heuristics that are hard to implement efficiently in Hadoop. There is also a study on finding the top-$k$ largest valued items in MapReduce [34], where each item has a *single total score*, which is clearly different from (and does not help) our case.

## 7. CLOSING REMARKS

Massive data is increasingly being stored and processed in MapReduce clusters, and this paper studies how to summarize this massive data using wavelet histograms. We designed both exact and approximation methods in MapReduce, which significantly outperform the straightforward adaptations of existing methods to MapReduce. Our methods are also easy to implement, in particular the two-level sampling method, making them appealing in practice.

Data summarization is an important technique for analyzing large datasets. The wavelet histogram is merely one representative, and there are many other types of summaries we may consider in the MapReduce model, such as other kinds of histograms (e.g., the V-optimal histogram [24]), various sketches and synopses, geometric summaries like $\varepsilon$-approximations and coresets, and graph summaries like distance oracles. Another open problem is how to incrementally maintain the summary when the data stored in the MapReduce cluster is being updated. Finally, data summarization in MapReduce is also an intellectually challenging problem, requiring a good blend of algorithmic techniques and system building.

## 8. ACKNOWLEDGEMENT

Jeffrey Jestes and Feifei Li were supported in part by NSF Grants IIS-0916488 and IIS-1053979. Ke Yi was supported by a Google Faculty Research Award.

# APPENDIX

## A. SYSTEM ISSUES OF H-WTOPK

**MapReduce Round 1:** In Round 1, a Mapper first computes local frequency vector $\mathbf{v}_j$ for split $j$ by using a hashmap to aggregate the total count for each key encountered as the records in the split are scanned. After $\mathbf{v}_j$ is constructed, we compute its wavelet coefficients in the Close interface of the Mapper. Since the number of nonzero entries in $\mathbf{v}_j$, denoted as $|\mathbf{v}_j|$, is typically much smaller than $u$, instead of running the $O(u)$-time algorithm of [26], we use the $O(|\mathbf{v}_j|\log u)$-time and $O(\log u)$-memory algorithm of [20]. During the computation we also keep two priority queues to store the top-$k$ and bottom-$k$ wavelet coefficients.

After all coefficients for split $j$ have been computed, the Mapper emits an intermediate key-value pair $(i, (j, w_{i,j}))$ for each of the top-$k$ and bottom-$k$ wavelet coefficients $w_{i,j}$ of the split. In the emitted pairs, the Mapper marks the $k$-th highest and the $k$-th lowest coefficients using $(i, (j+m, w_{i,j}))$ and $(i, (j+2m, w_{i,j}))$, respectively. Finally, the Mapper saves all the unemitted coefficients as state information to an HDFS file associated with split $j$.

After the Map phase, the Reducer receives the top-$k$ and bottom-$k$ wavelet coefficients from all the splits, $2km$ of them in total. We denote by $R$ the set of distinct indices of the received coefficients. For each index $i \in R$, The Reducer passes the corresponding $(j, w_{i,j})$'s received to a Reduce function, which adds up these $w_{i,j}$'s, forming a partial sum $\widehat{w}_i$ for $w_i$. Meanwhile we construct a bit vector $F_i$ of size $m$ such that $F_i(j) = 0$ if $w_{i,j}$ has been received and $F_i(j) = 1$ if not. While examining the $(j, w_{i,j})$'s in the Reduce function, if we encounter a marked pair, we remember it so the $k$-th highest and the $k$-th lowest coefficient from each split $j$, denoted as $\tilde{w}_j^+$ and $\tilde{w}_j^-$, can be obtained.

After we have all partial sums $\widehat{w}_i$ for all $i \in R$, and $\tilde{w}_j^+, \tilde{w}_j^-$ for all $j$, we compute upper bound $\tau_i^+$ (resp. lower bound $\tau_i^-$) on $w_i$, by adding $\sum_{j=1}^{m} F_i(j)\tilde{w}_j^+$ (resp. $\sum_{j=1}^{m} F_i(j)\tilde{w}_j^-$) to $\widehat{w}_i$. Then we obtain a lower bound $\tau_i$ on $|w_i|$, hence $T_1$, as described in Section 3. Finally, we save tuple $(i, \widehat{w}_i, F_i)$ for all $i \in R$, and $T_1$ as state information in a local file on the Reducer machine.

**MapReduce Round 2:** To start Round 2, $T_1/m$ is first set as a variable in the Job Configuration. In this round, for the Map phase we define an alternate InputFormat so a Mapper does not read an input split at all. Instead, a Mapper simply reads state information, i.e., all wavelet coefficients not sent in Round 1, one by one. For any $w_{i,j}$ such that $|w_{i,j}| > T_1/m$, a Mapper emits the pair $(i, (j, w_{i,j}))$.

The Reducer first reads tuple $(i, \widehat{w}_i, F_i)$ for all $i \in R$ from the local file written in Round 1. For each $i$, it passes all corresponding $(j, w_{i,j})$'s received in this round to a Reduce function. Now we update partial sum $\widehat{w}_i$ by adding these new coefficients, and update $F_i$ correspondingly. We also refine upper bound $\tau_i^+$ (resp. lower bound $\tau_i^-$) as $\tau_i^+ = \widehat{w}_i + \|F_i\|_1 \cdot T_1/m$ (resp. $\tau_i^- = \widehat{w}_i - \|F_i\|_1 \cdot T_1/m$), where $\|F_i\|_1$ denotes the number of 1's in $F_i$.

With the updated $\tau_i^+, \tau_i^-$, we obtain a new $T_2$, which can be used to prune indices from $R$ as described in Section 3. Lastly, the Reducer writes updated $\widehat{w}_i$ for all $i \in R$ in a local file, and the set of candidate indices $R$ in an HDFS file.

**MapReduce Round 3:** In Round 3, the master reads $R$ from HDFS and adds it to the Distributed Cache. Like in Round 2, the Mappers still do not read the input splits. During initialization, each Mapper reads $R$ from the distributed cache. Then, it reads from the state file storing the wavelet coefficients. For any $w_{i,j}$ it checks if $i \in R$ and $|w_{i,j}| \leq T_1/m$. If so, it means it has not been communicated to the Reducer yet, and thus we emit $(i, (j, w_{i,j}))$.

On the Reducer side, similar to Round 2, the Reducer first reads $R$ and $\widehat{w}_i$ for all $i \in R$'s from the local file. Then for each $i$, the Reduce function adds all newly received $w_{i,j}$'s to $\widehat{w}_i$, yielding accurate $w_i$. Finally, we return the top-$k$ coefficients $w_i$ of largest magnitude for $i \in R$ as the best $k$-term representation for $\mathbf{v}$.

## B. SYSTEM ISSUES OF TWOLEVEL-S

As before, we will have $m$ Mappers, one per input split, and 1 Reducer. The first issue is how to randomly read records from an input split. The default Hadoop RecordReader in InputFile format is designed to sequentially scan an input split. Hence, we define our own InputFile format *RandomInputFile*, assuming each record in the input files has a fixed size. The RandomInputFile defines a custom RecordReader, called RandomRecordReader, which can randomly sample records from an input split. A straightforward implementation is to simply seek to a random offset in the split when the Mapper requests the next record, but this requires seeking offset locations in both directions. Instead, we implement it as follows.

When the RandomRecordReader is first initialized, it determines $n_j$, the number of records in the split. Next, it randomly selects $pn_j$ offsets in the split, where $p = 1/(\varepsilon^2 n)$ is the sample probability of the first-level sampling, and stores them in a priority queue $Q$ sorted in ascending order. Afterwords, every time the RandomRecordReader is invoked by the Mapper to retrieve the next record from the split, it seeks to the record indicated by the next offset, and retrieves the record there. We continue this process iteratively until all $pn_j$ random records have been obtained. Note in Section 4, we assume coin-flip sampling for sake of simpler analysis; here we use sampling without replacement. It has been observed coin-flip sampling and sampling without replacement behave almost the same for most sampling-based methods [37], and we observe this is also true for our sampling-based approaches.

Using RandomInputFile as the InputFile format, two-level sampling can be implemented in one round of MapReduce, as follows.

**Map phase.** During initialization of the Map phase we specify $n$ and $\varepsilon$ in the Job Configuration. With the RandomRecordReader, the MapRunner reads the $pn_j$ random records one at a time and invokes the Map function for each record, which simply maintains aggregated counts for keys of the sampled records. After the MapRunner has processed all sampled records, the Mapper's Close routine is called. It iterates over all sampled keys and checks their aggregate counts. If $\mathbf{s}_j(x) \geq 1/(\varepsilon\sqrt{m})$, we emit the pair $(x, \mathbf{s}_j(x))$. Otherwise, we emit $(x, 0)$ with probability $\varepsilon\sqrt{m} \cdot s_j(x)$.

**Reduce phase.** For each key $x$, the Reducer passes all corresponding $(x, \mathbf{s}_j(x))$ or $(x, 0)$ pairs to the Reduce function, which computes the estimated $\widehat{\mathbf{v}}(x)$ as described in Section 4. After all keys are processed by the Reducer, its Close method is invoked, where approximate wavelet coefficients are computed from approximate global frequency vector $\widehat{\mathbf{v}}$. In the end, we emit $(i, w_i)$ pairs for the top-$k$ approximate coefficients (with the $k$ largest magnitudes).

**Remarks.** In our discussion so far, our RandomRecordReader assumes fixed length records. However, it is easy to extend it to support variable length records as well. Instead, assume records of variable length end with a 4-byte record length followed by a delimiter character or byte sequence (e.g., a new line character). The RandomRecordReader initially generates $pn_j$ random offsets and inserts them in a priority queue $Q$. It then processes offsets from $Q$ one at a time. It seeks to an offset and scans forward until it finds the record length $r$ and delimiter (this is easy to achieve with a few-bytes look-ahead buffer). The RandomRecordReader determines the start offset $o$ of the record, using $r$ and the end offset of the record, and records an $(o, r)$ pair in a Heap $H$ sorted by $o$. It is possible some of the $pn_j$ randomly selected offsets may point to the contents of the same record. Note RandomRecordReader processes sampled offset locations in (ascending) sorted order. Hence, whenever the RandomRecordReader processes an offset $o$, it determines if its current split position is larger than $o$. If it is, $o$ points to the contents of the same record as the last processed offset. In this case, RandomRecordReader randomly generates a new offset $o'$ (to replace $o$) which does not point to locations covered by any $(o, r)$ pairs in $H$; this is easy to ensure using the information in $H$. If $o'$ appears before the current split location of the RandomRecordReader, the end split offset is added to $o'$. Then, the RandomRecordReader pushes $o'$ onto the priority queue $Q$ (which contains yet to be processed offsets) and continues with the next offset. If the end of the split is reached, and $Q$ is not empty, the RandomRecordReader goes back to the head of the split to continue processing offsets in $Q$; first subtracting the end split offset from offsets in $Q$. The process is repeated until enough records are sampled ($Q$ becomes empty). At this point $H$ contains sorted offsets of all sampled $pn_j$ records (and their lengths). The RandomRecordReader seeks to and reads all records at offsets in $H$ by ascending order.

Another issue which must be addressed is how the RandomRecordReader should determine the number of records in the $j$th split $n_j$ when records are variable length, so that it may determine the number of records which should be sampled. We can either assume that this information is made available when the datasets were initially loaded into the MapReduce cluster, or statistics such as the average, median, or minimum record sizes can be used to determine an appropriate $n_j$ depending on the application. Additionally, in order to decrease the likeliness that any two of the $pn_j$ randomly selected offsets point to the contents of the same record, it can be enforced that the offsets do not appear within the average, median, or minimum record size of each other.

In the implementation of our exact and sampling methods, we choose to do the final processing in the close method, instead of using the combiner. This is a technicality due to the default behavior of Hadoop, which runs the COMBINE function continuously while keys are being processed to save memory buffer space (leads to fewer disk writes). On the other hand, the close method is guaranteed to run only once when all keys have been processed.